\documentclass[%
 reprint,
 superscriptaddress,
 showpacs,preprintnumbers,
 amsmath,amssymb,
 aps,
 prb,
 longbibliography,
]{revtex4-1}

\usepackage{graphicx}
\usepackage{dcolumn}
\usepackage{bm}
\usepackage{color}
\usepackage{ulem}

\begin{document}

\preprint{APS/123-QED}

\title{
Thermally-induced Phases in an Ising Kondo Lattice Model on a Triangular Lattice:
Partial Disorder and Kosterlitz-Thouless State
}

\author{Hiroaki Ishizuka}
\affiliation{
Department of Applied Physics, University of Tokyo, Tokyo 113-8656, Japan
}

\author{Yukitoshi Motome}%
\affiliation{
Department of Applied Physics, University of Tokyo, Tokyo 113-8656, Japan
}

\date{\today}

\begin{abstract}
Magnetic and electronic properties of a Kondo lattice model with Ising localized spins are studied on an isotropic triangular lattice.
By using Monte Carlo simulation, we present that the model shows a rich phase diagram with four dominant states: two-sublattice stripe, three-sublattice ferrimganetic, partially disordered, and Kosterlitz-Thouless like quasi-long-range ordered states. 
Among them, the partially disordered state and Kosterlitz-Thouless like state are intermediate phases induced by thermal fluctuations in the phase competing regime; they are present only at finite temperatures and eventually taken over by another phases as the temperature is further lowered.
Although the Kosterlitz-Thouless like state was found also in triangular Ising antiferromagnets with further-neighbor interactions, the partially disordered state has not been reported in the localized spin only models in two dimensions.
Interestingly, the partially disordered phase is also peculiar in the charge degree of freedom of itinerant electrons; it is insulating and accompanied by charge disproportionation.
From a combined analysis of a mean-field calculation of the band structure and Monte Carlo simulation, we conclude that the partial disorder in the present model is stabilized by the Slater mechanism.
\end{abstract}


\maketitle

\section{\label{sec:intro}
Introduction
}

\begin{figure}
   \includegraphics[width=0.8\linewidth]{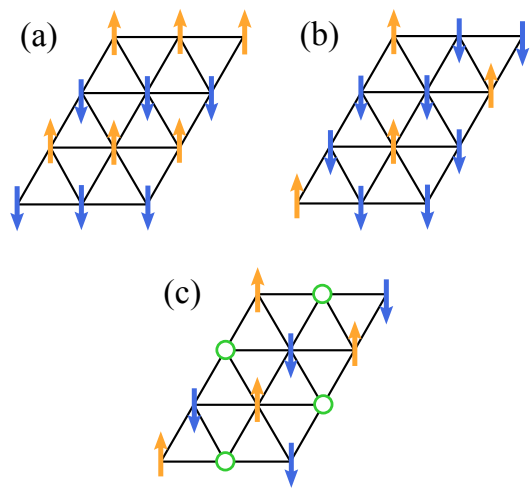}
   \caption{(Color online).
   Schematic pictures of (a) stripe order, (b) ferrimagnetic (FR) order, and (c) partial disorder (PD) on a triangular lattice.
   The arrows show magnetically ordered sites and the open circles are thermally fluctuating paramagnetic sites.
   }
   \label{fig:model}
\end{figure}

The antiferromagnetic (AF) Ising model on a triangular lattice is one of the most fundamental models for geometrically frustrated systems. 
When the interaction is restricted to the nearest-neighbor (NN) pairs, frustration in each triangle prevents the system from forming a long-range order (LRO) down to zero temperature, and the ground state has extensive degeneracy and associated residual entropy~\cite{Wannier1950,Houtappel1950,Husimi1950}.
The degenerate ground state is extremely sensitive to perturbations.
For instance, an infinitesimal second-neighbor interaction lifts the degeneracy and induces a LRO in the ground state; a two-sublattice stripe order [Fig.~\ref{fig:model}(a)] is selected as the ground state when the additional interaction is AF, while a three-sublattice ferrimagnetic (FR) order [Fig.~\ref{fig:model}(b)] is selected for the ferromagnetic (FM) interaction.

In such a degenerate situation, thermal fluctuations also play an interesting role. 
In general, there is a possibility that a high-entropic state is selected out of the ground state manifold by raising temperature ---this is called the order by disorder~\cite{Villain1977}. 
For the AF Ising model, a candidate for such an emergent state is a partially disordered (PD) state. 
The PD state is peculiar coexistence of magnetically ordered moments and thermally-fluctuating paramagnetic moments.
Such possibility was first discussed by the mean-field study in the presence of second-neighbor FM interaction~\cite{Mekata1977}; the mean-field study predicted that a three-sublattice PD phase with an AF ordering on the honeycomb subnetwork and paramagnetic moments at the remaining sites [Fig.~\ref{fig:model}(c)] was induced at finite temperature from the degenerate manifold in the limit of vanishing second-neighbor interaction. 
Although such PD state was experimentally observed in several Co compounds~\cite{Kohmoto1998,Niitaka2001} and theoretically shown to present in a stacked triangular lattice model~\cite{Todoroki2004}, Monte Carlo (MC) simulations in two-dimensional triangular lattice models have indicated that PD is fragile and remains at most as a quasi-LRO; namely, in most cases, the PD state is taken over by another peculiar intermediate state, the Kosterlitz-Thouless (KT) state~\cite{Wada1982,Fujiki1983,Landau1983,Takayama1983,Fujiki1984,Takagi1995}.

On the other hand, recently, the authors have studied Ising-spin Kondo lattice models on a triangular lattice~\cite{Ishizuka2012} and kagome lattice~\cite{Ishizuka2012-3} by MC simulation, and showed the presence of PD state in the purely two-dimensional models.
In these models, the interplay between localized moments and itinerant electrons plays a crucial role in the following points. 
First, the kinetic motion of electrons induces effective interactions known as the Ruderman-Kittel-Kasuya-Yosida (RKKY) mechanism~\cite{Ruderman1954,Kasuya1956,Yosida1957}. 
The long-ranged and oscillating nature of the interactions drives keen competition between different magnetic states.
Furthermore, the change of magnetic states affects the electronic state in a self-consistent manner through the spin-charge coupling; the system can gain the energy by forming some particular electronic state associated with magnetic ordering.  
In the previous study,  the authors suggested that the PD state is stabilized by the non-perturbative role of itinerant electrons~\cite{Ishizuka2012}.

In this contribution, we present our comprehensive numerical results on the magnetic and electronic properties of the Ising-spin Kondo lattice model on a triangular lattice. 
To further clarify the stabilization mechanism of PD, we analyze the evolution of band structure under the PD type magnetic texture on the basis of a simple mean-field argument. 
The analysis suggests that the spin-charge coupling can stabilize the PD state by the Slater mechanism.
Bearing this mean-field picture in mind, we present and discuss the results of MC simulation in details. 
We distinguish the two intermediate-temperature states, PD and KT-like states, from the two-sublattice stripe and three-sublattice FR LRO states, and identify the range of the phases by varying the electron filling and the strength of spin-charge coupling. 
Analyzing the phase diagram and electronic states in comparison with the mean-field picture, we conclude that the two-dimensional PD state is stabilized through the Slater mechanism.

The organization of this paper is as follows.
In Sec.~\ref{sec:model_and_method}, we introduce the model and method.
The definitions of physical quantities we calculated are also given.
In Sec.~\ref{sec:mft}, we present the mean-field analyses on the band structure in the PD state.
MC results are presented for magnetic properties in Sec.~\ref{sec:mc} and for electronic properties in Sec.~\ref{sec:estruct}.
Section~\ref{sec:summary} is devoted to summary.

\section{\label{sec:model_and_method}
Model and Method
}

In this section, we introduce the model and method.
The model is given in Sec.~\ref{sec:model} and the MC method is described in Sec.~\ref{sec:method}.
In Sec.~\ref{sec:pmoment}, we give the definitions of physical quantities that we used to elaborate the phase diagram and thermodynamic properties.

\subsection{
Model
\label{sec:model}
}

We consider a single-band Kondo lattice model on a triangular lattice with localized Ising spin moments.
The Hamiltonian is given by
\begin{eqnarray}
H = -t \! \sum_{\langle i,j \rangle, \sigma} \! ( c^\dagger_{i\sigma} c_{j\sigma} + \text{H.c.} ) + J \sum_{i}\sigma_i^z S_i.
\label{eq:H}
\end{eqnarray}
The first term represents hopping of itinerant electrons, where $c_{i\sigma}$ ($c^\dagger_{i\sigma}$) is
the annihilation (creation) operator of an itinerant electron with spin $\sigma= \uparrow, \downarrow$ at
$i$th site, and $t$ is the transfer integral.
The sum $\langle i,j \rangle$ is taken over nearest-neighbor (NN) sites on the triangular lattice.
The second term is the onsite interaction between localized spins and itinerant electrons, where $\sigma_i^z = c_{i\uparrow}^\dagger c_{i\uparrow} -  c_{i\downarrow}^\dagger c_{i\downarrow}$ represents the $z$-component of itinerant electron spin, and $S_i = \pm 1$ denotes the localized Ising spin at $i$th site; $J$ is the coupling constant (the sign of $J$ does not matter in the present model). 
Hereafter, we take $t=1$ as the unit of energy, the lattice constant $a = 1$, and the Boltzmann constant $k_{\rm B} = 1$.

\subsection{
Monte Carlo simulation
\label{sec:method}
}

To investigate thermodynamic properties of the model (\ref{eq:H}), we adopted a MC simulation which is widely used for similar models~\cite{Yunoki1998}.
The model belongs to the class of models in which fermions are coupled to classical fields.
For this class of models, the partition function is given by
\begin{eqnarray}
Z={\rm Tr}_f{\rm Tr}_c \exp[\beta(H-\mu\hat{N_e})],
\end{eqnarray}
where $\beta=1/T$ is the inverse temperature, $\mu$ is the chemical potential, and $\hat{N_e}$ is the total number operator for fermions.
Here, ${\rm Tr}_f$ is the trace over classical degree of freedom (in the current case, Ising spin configurations), and ${\rm Tr}_c$ is the trace over itinerant fermions. 
In the MC simulation, ${\rm Tr}_f$ is calculated by using the Markov-chain MC sampling. 
MC updates are done by the usual single-spin flip on the basis of the standard METROPOLIS algorithm. 
The MC weight is calculated by taking the fermion trace ${\rm Tr}_c$ for each configuration of classical variables in the following form, 
\begin{eqnarray}
P(\{S_i \}) = \exp[ -S_{\rm eff}(\{S_i \}) ],
\label{eq:P}
\end{eqnarray}
where $S_{\rm eff}$ is the effective action calculated as
\begin{eqnarray}
S_{\rm eff}(\{S_i \}) = - \sum_\nu \log[1 + \exp\{-\beta(E_\nu(\{S_i \})-\mu)\}].
\label{eq:S_eff}
\end{eqnarray}
Here, $E_\nu(\{S_i \})$ are the energy eigenvalues for the configuration $\{S_i \}$, which are readily calculated by the exact diagonalization as it is a one-particle problem in a static potential.

The calculations were conducted for the system sizes $N=12 \times 12$, $15 \times 15$, $12 \times 18$, and $18 \times 18$ under the periodic boundary conditions.
Thermal averages of physical quantities were calculated for typically 4300-9800 MC steps after 1700-5000 steps for thermalization. 
The results are shown in the temperature range where the acceptance ratio is roughly larger than 1\%.
We divide the MC measurements into five bins and estimate the statistical errors by the standard deviations among the bins.

\subsection{
Physical quantities
\label{sec:pmoment}
}

As we will see later, the model (\ref{eq:H}) exhibits phase transitions to various magnetic states including different types of three-sublattice orders: ferrimagnetic (FR) state [Fig.~\ref{fig:model}(b)] and partially disordered (PD) state [Fig.~\ref{fig:model}(c)].
These magnetic states, in principle, are distinguishable by the spin structure factor for the Ising spins,
\begin{eqnarray}
S({\bf q}) = \frac{1}{N} \sum_{i,j} \langle S_i S_j \rangle \exp({\rm i} {\bf q}\cdot{\bf r}_{ij}),
\label{eq:Sq}
\end{eqnarray}
where the braket denotes the thermal average in the grand canonical ensemble, and ${\bf r}_{ij}$ is the position vector from $i$ to $j$th site. 
The PD order is signaled by peaks of $S({\bf q})$ at ${\bf q}=\pm(2\pi/3,-2\pi/3)$, while the FR order develops a peak at ${\bf q}=0$ in addition to ${\bf q}=\pm(2\pi/3,-2\pi/3)$.
No Bragg peaks develop in the KT state as it is a quasi-LRO.
However, in finite-size calculations, it is difficult to distinguish these phases solely by the structure factor, as the correlation length in the KT state is divergent and easily exceeds the system size at low temperature.

For distinguishing the FR, PD, and KT instabilities, it is helpful to use the pseudospin defined for each three-site unit cell:
\begin{eqnarray}
\tilde{\bf S}_m = 
\left(
\begin{array}{ccc}
\frac2{\sqrt6} & -\frac1{\sqrt6} & -\frac1{\sqrt6} \\
0              &  \frac1{\sqrt2} & -\frac1{\sqrt2} \\
\frac1{\sqrt3} &  \frac1{\sqrt3} &  \frac1{\sqrt3} \\
\end{array}
\right)
\left(
\begin{array}{c}
S_i  \\
S_j  \\
S_k  \\
\end{array}
\right),
\end{eqnarray}
and its summation 
\begin{eqnarray}
\tilde{\bf M} = \frac{3}{N} \sum_m  \tilde{\bf S}_m 
\end{eqnarray}
where $m$ is the index for the three-site unit cells, and $(i,j,k)$ denote the three sites in the $m$th unit cell belonging to the sublattices (A,B,C), respectively~\cite{Takayama1983,Fujiki1984}.
Then, the three-sublattice PD state [Fig.~\ref{fig:model}(c)] is characterized by a finite $\tilde{\bf M} = (\tilde{M}_x,\tilde{M}_y,\tilde{M}_z)$ parallel to $(\sqrt{3/2},1/\sqrt2,0)$, $(0,\sqrt2,0)$, or their threefold symmetric directions around the $z$-axis.
On the other hand, the three-sublattice FR state [Fig.~\ref{fig:model}(b)] is characterized by a finite $\tilde{\bf M}$ along $(\sqrt{2/3},\sqrt2,1/\sqrt3)$, $(2\sqrt{2/3},0,-1/\sqrt3)$, or their threefold symmetric directions around the $z$-axis.
Hence, the two states are distinguished by the azimuth of $\tilde{\bf M}$ in the $xy$-plane as well as $M_z$.
In the MC calculations, we measure 
\begin{eqnarray}
M_{xy} &=& \langle (\tilde{M}_x^2 + \tilde{M}_y^2)^{1/2} \rangle, 
\label{eq:Mxy} \\
M_z &=& \langle |\tilde{M}_z| \rangle,
\label{eq:Mz}
\end{eqnarray}
and the corresponding susceptibilities,
\begin{eqnarray}
\chi_{xy} &=& \frac{N}{T} (\langle \tilde{M}_x^2 + \tilde{M}_y^2 \rangle - M_{xy}^2 ), \\
\chi_z &=& \frac{N}{T} (\langle \tilde{M}_z^2 \rangle - M_z^2 ).
\end{eqnarray}
We also introduce the azimuth parameter of $\tilde{{\bf M}}$ defined by
\begin{eqnarray}
\psi = {\cal M}^3 
\cos{6 \phi_M},
\label{eq:psi}
\end{eqnarray}
where $\phi_M$ is the azimuth of $\tilde{\bf M}$ in the $xy$ plane and ${\cal M} = \frac38 M_{xy}^2$.
The parameter $\psi$ has a negative value and $\psi \to -\frac{27}{64}$ for the perfect PD ordering, while it becomes positive and $\psi \to 1$ for the perfect FR ordering; $\psi=0$ for both paramagnetic and KT phases in the thermodynamic limit $N \to \infty$.

In addition, we calculate the spin entropy to distinguish the three-sublattice orderings.
The spin entropy per site is defined by
\begin{eqnarray}
{\cal S}(T) = -\frac{1}{N} \sum_{\{S_i\}} P(\{S_i\})\log P(\{S_i\}),
\label{eq:Sdef}
\end{eqnarray}
where $P(\{S_i\})$ is the probability for spin configuration $\{S_i\}$ to be realized, given in Eq.~(\ref{eq:P}).
In the actual MC calculation, instead of directly calculating Eq.~(\ref{eq:Sdef}), ${\cal S}$ is evaluated by calculating its temperature derivative
\begin{eqnarray}
\frac{\partial{\cal S}(T)}{\partial T} = \frac{1}{NT^2} \left\{ \langle S_{\rm eff} H \rangle - \langle S_{\rm eff}\rangle \langle H \rangle \right\},
\label{eq:delSdelT}
\end{eqnarray}
and integrating it as 
\begin{equation}
{\cal S}(T) = \int_0^T \frac{\partial{\cal S}(T)}{\partial T} dT = \log 2 - \int_T^\infty \frac{\partial{\cal S}(T)}{\partial T} dT. 
\label{eq:Sint}
\end{equation}
In Eq.~(\ref{eq:delSdelT}), $S_{\rm eff}$ is the effective action in Eq.~(\ref{eq:S_eff}). 
In the following calculations, we set the cutoff $T=1$ for the upper limit of the last integral in Eq.~(\ref{eq:Sint}).

On the other hand, in order to identify the two-sublattice stripe order [Fig.~\ref{fig:model}(a)], we calculate the order parameter
\begin{eqnarray}
M_{{\rm str}} = \left[ \sum_{{\bf q}^{*}_{\rm str}} \left\{ \frac{S({\bf q}^{*}_{\rm str})}{N} \right\}^2 \right]^{1/2},
\label{eq:Mstr}
\end{eqnarray}
and its susceptibility $\chi_{\rm str}$. 
Here, the sum is taken for the characteristic wave vectors of the stripe orders running in three different directions, ${\bf q}^{*}_{\rm str}= (\pi,0)$ and $(\pm\frac12\pi,\frac{\sqrt3}2\pi)$.

We also examine the thermodynamic behavior of electronic states for itinerant electrons.
There, we computed the charge modulation defined by
\begin{eqnarray}
n_{\rm CO} = \left\{\frac{N({\bf q}^*_{\rm CO})}{N}\right\}^{1/2}
\label{eq:n_CO}
\end{eqnarray}
at ${\bf q}^*_{\rm CO}=(-2\pi/3,2\pi/\sqrt3)$, which corresponds to the wave numbers for the three-sublattice orders.
Here, $N({\bf q})$ is the charge structure factor for itinerant electrons,
\begin{eqnarray}
N({\bf q}) = \frac{1}{N} \sum_{i,j} \langle n_i n_j \rangle \exp({\rm i} {\bf q}\cdot{\bf r}_{ij}),
\end{eqnarray}
where $n_i = \frac12\sum_\sigma c_{i\sigma}^\dagger c_{i\sigma}$.

\section{
Mean-field band structure
\label{sec:mft}
}

Before going to the MC results, we here discuss how one particle band structure is modulated by PD ordering in a mean-field picture. 
We consider a three-sublattice LRO state, in which the localized spins give a mean-field local magnetic field to itinerant electrons.
Namely, we consider a mean-field Hamiltonian given by
\begin{eqnarray}
{\cal H}^{\rm MF} = \sum_{\bf k}
\begin{pmatrix}
\Delta_{\mathrm{A}}\sigma^z_\alpha & \tau_{\bf k} & \tau_{\bf k}^\ast \\
 \tau_{\bf k}^\ast & \Delta_{\mathrm{B}}\sigma^z_\alpha & \tau_{\bf k} \\
 \tau_{\bf k} & \tau_{\bf k}^\ast & \Delta_{\mathrm{C}}\sigma^z_\alpha
\end{pmatrix}\label{eq:mfh}
.
\end{eqnarray}
Here, three rows correspond to the different sublattices A, B, and C in the three-site unit cell; $\Delta_\alpha$ is a mean field given by $J\langle S_\alpha\rangle$ ($\alpha=\mathrm{A}, \mathrm{B}, \mathrm{C}$).
The sum is taken in the first Brillouin zone for the magnetic unit cell for three-sublattice order.
$\tau_{\bf k}$ is the hopping term for itinerant electrons given by
\begin{eqnarray}
\tau_{\bf k} = -t[e^{{\rm i}k_x} + e^{{\rm i}\left(-\frac{k_x}2 + \frac{\sqrt3}2 k_y\right)} + e^{{\rm i}\left(-\frac{k_x}2 - \frac{\sqrt3}2 k_y\right)}]
\end{eqnarray}
and $\sigma^z_\alpha$ corresponds to the $z$ component of itinerant electron spin in each sublattice $\alpha$.

The band structure for a FR order, $(\Delta_{\rm A},\Delta_{\rm B},\Delta_{\rm C})=( \Delta, \Delta, -\Delta)$, was recently studied by the authors~\cite{Ishizuka2012-2}.
There, it was reported that the electronic structure in the FR order is semimetallic with forming Dirac nodes at the electron filling $n= \frac{1}{2N}\sum_{i\sigma} \langle c_{i\sigma}^\dagger c_{i\sigma} \rangle=1/3$ for $J>t$.

\begin{figure}
   \includegraphics[width=0.92\linewidth]{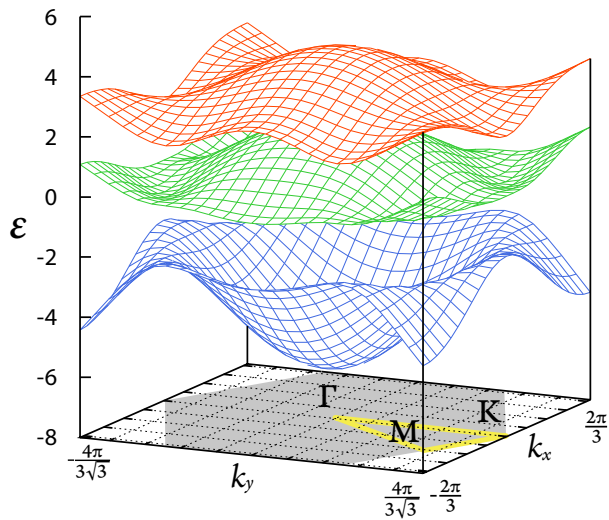}
   \caption{(Color online).
   Mean-field band structure calculated by Eq.~(\ref{eq:mfh}) for the local magnetic field of PD type, $(\Delta_{{\rm A}},\Delta_{{\rm B}},\Delta_{{\rm C}})=(2,0,-2)$.
   Each of the three bands shown is doubly degenerate, and there are totally six bands.
   The gray hexagon on the basal plane shows the first Brillouin zone for the magnetic supercell.
   }
   \label{fig:band1}
\end{figure}

\begin{figure}
   \includegraphics[width=0.8\linewidth]{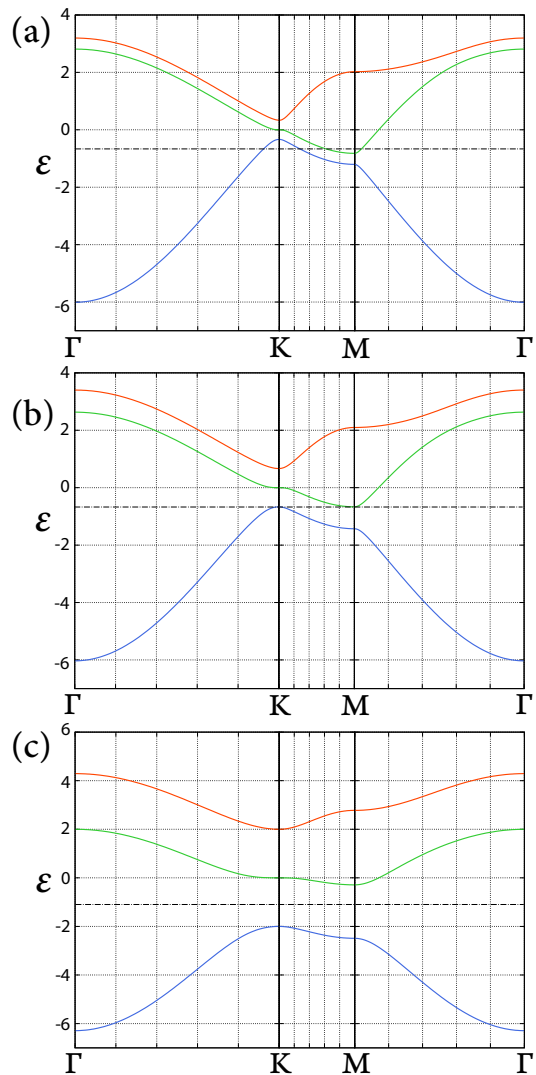}
   \caption{(Color online).
   Mean-field band structure along the symmetric lines in the local magnetic field of PD type, $(\Delta_{\rm A},\Delta_{\rm B},\Delta_{\rm C})=(\Delta,0,-\Delta)$: (a) $\Delta=1/3$, (b) $\Delta=2/3$, and (c) $\Delta=2$.
   The dashed horizontal lines indicate the Fermi level for $n=1/3$.
   }
   \label{fig:band2}
\end{figure}

Here, we discuss the band structure for the PD case, $(\Delta_{\rm A},\Delta_{\rm B},\Delta_{\rm C})=(\Delta,0,-\Delta)$.
The band structure for $\Delta=2$ is shown in Fig.~\ref{fig:band1}.
In this case, all three bands shown in the figure are doubly degenerate and there are six bands in total.
The first Brillouin zone is shown by the gray shade in the bottom surface. 
The result shows the presence of an energy gap at the Fermi level corresponding to $n=1/3$, that opens between the lowest energy band and the middle band [see also Fig.~\ref{fig:band2}(c)].

We next look into the conditions for the energy gap formation in the mean-field PD band.
Figure~\ref{fig:band2} shows the results of band structure while varying $\Delta$.
The results are plotted along the symmetric line in the Brillouin zone shown in the bottom surface in Fig.~\ref{fig:band1}.
For small $\Delta$, the system is metallic at $n=1/3$, as shown in the case of $\Delta=1/3$ in Fig.~\ref{fig:band2}(a); both electron and hole pockets are present at the Fermi level.
The pockets shrink as increasing $\Delta$, and disappear at the same time at $\Delta=2/3$, as shown in Fig.~\ref{fig:band2}(b).
For larger $\Delta$, an energy gap opens between the lowest and middle bands, corresponding to $n=1/3$, as stated above [Fig.~\ref{fig:band2}(c)]. 
Hence, $\Delta_c=2/3$ is the critical point for the metal-insulator transition in this mean-field PD state.

\begin{figure}
   \includegraphics[width=0.9\linewidth]{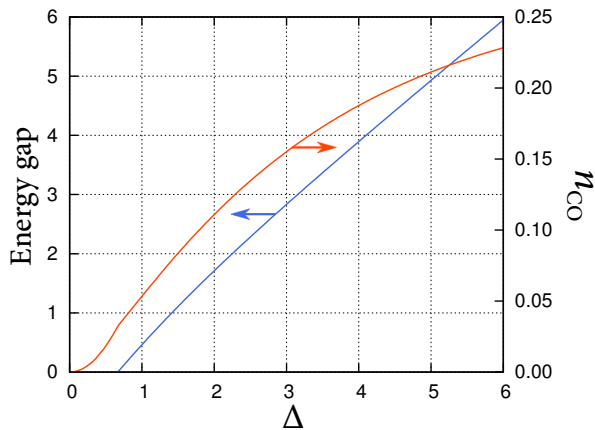}
   \caption{(Color online).
   $\Delta$ dependences of the mean-field energy gap and associated charge modulation $n_{\rm CO}$ at $n=1/3$.
   }
   \label{fig:gap}
\end{figure}

Figure~\ref{fig:gap} shows $\Delta$ depedences of the energy gap and associated charge modulation $n_{\rm CO}$ [Eq.~(\ref{eq:n_CO})] at $n=1/3$.
The charge gap develops for $\Delta > 2/3$ and monotonically increases, approaching asymptotically a $\Delta$-linear form as $\Delta \gg t$.
The charge modulation is induced by the inhomogeneity of local potential; the local charge density at B sites (the site corresponds to paramagnetic sites) becomes dilute compared to those at A and C sites (the magnetically ordered sites).
In the limit of $\Delta \gg t$, $n_{\rm CO}$ approaches $n_{\rm CO}=1/\sqrt{12}\sim 0.289$.

The results above suggest a stabilization mechanism of PD which is absent in the localized spin only model.
In the previous studies on the Ising spin models~\cite{Takayama1983,Fujiki1983,Wada1982} and an equivalent classical particle model~\cite{Landau1983} on a triangular lattice, PD was shown to be unstable against thermal fluctuations and taken over by a KT state.
In the case of our model, however, as the KT state lacks a long-range periodic magnetic structure, it is expected that the KT state does not open an energy gap in the electronic state of itinerant electrons.
Therefore, in contrast to the case of localized spin only models, there is a chance for the current model to stabilize the PD state by the Slater mechanism, that is, by forming an energy gap at the Fermi level with folding the Brillouin zone under a periodic magnetic order.

In addition, the formation of an energy gap for $\Delta > 2/3$ implies that, if the PD state is stabilized by the Slater mechanism, it should appear from a finite $J$, and not remain stable down to $J\to0$.
This is in sharp contrast to magnetic ordering by the Ruderman-Kittel-Kasuya-Yosida (RKKY) interaction~\cite{Ruderman1954,Kasuya1956,Yosida1957}; as the RKKY interaction is given by the second-order perturbation in terms of $J/t$, if the PD state is stabilized by the RKKY interaction, it should appear for an infinitesimal $J$.
Hence, the phase diagram in the small $J$ region gives an idea on how the PD state is stabilized.
We will discuss this point by showing the MC results while changing $J$ in the next section.

\section{
Monte Carlo simulation
\label{sec:mc}
}
In this section, we present the results of MC simulation introduced in Sec.~\ref{sec:method}.
We first show the finite-temperature phase diagrams in Sec.~\ref{sec:pdiag}, which include four magnetic phases: stripe, PD, FR, and KT-like states.
The details of numerical data for the PD state are elaborated in Sec.~\ref{sec:pd}.
The results for stripe, KT-like, and FR states are discussed  in Sec.~\ref{sec:stripe_ferri}.

\subsection{
Phase diagrams
\label{sec:pdiag}
}

\begin{figure}
   \includegraphics[width=0.8\linewidth]{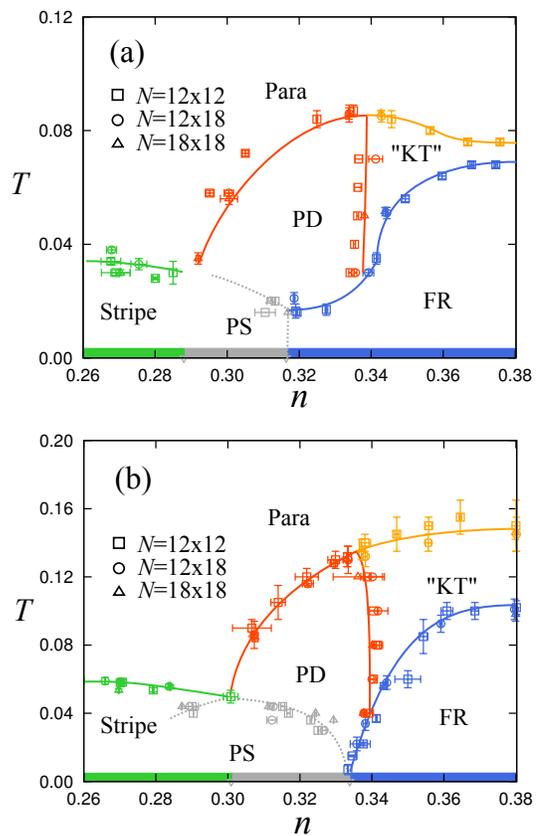}
   \caption{(Color online).
   Phase diagrams of the model~(\ref{eq:H}) while varying $n$ at (a) $J=1$ and (b) $J=2$.
   The symbols show phase boundaries for the four phases: stripe, partially disordered (PD), KT-like (``KT"), and ferrimagnetic (FR) phases.
   PS represents a phase separation. The lines are guides for the eyes.
   The strips at $T=0$ show the ground states obtained by comparing the energy of stripe and FR states.
   }
   \label{fig:ndiag}
\end{figure}

Figure~\ref{fig:ndiag}(a) shows the phase diagram around the electron filling $n= 1/3$ at $J=1$ obtained by MC calculations. 
There are four dominant ordered phases ---stripe, FR, PD, and KT-like phases, in addition to an electronic phase separation (PS).
The strip at the bottom of the figure shows the ground state obtained by variational calculation comparing the ground state energy of the stripe and FR states (the details of variational calculation is given in Appendix~\ref{sec:pseparation}).
For the relatively low filling of $n \lesssim 0.29$, the stripe order with period two [Fig.~\ref{fig:model}(a)] develops in the low temperature region.
On the other hand, for the higher filling of $n \gtrsim 0.32$, the system exhibits the three-sublattice FR order at low temperature [Fig.~\ref{fig:model}(b)]. 
MC data for the stripe and FR orders will be discussed in Sec.~\ref{sec:stripe_ferri}.
In addition to these two states, the numerical results show two intermediate-temperature states depending on the electron filling $n$.
For $0.29 \lesssim n \lesssim 0.34$, we identify the intermediate phase as the three-sublattice PD state [Fig.~\ref{fig:model}(c)]. 
The details will be discussed in Sec.~\ref{sec:pd}.
Meanwhile, for $n \gtrsim 0.34$, we find KT-like behavior similar to the one discussed in the Ising models~\cite{Takayama1983,Fujiki1984,Wada1982,Fujiki1983,Landau1983}, as presented in Sec.~\ref{sec:stripe_ferri}.
In these intermediate-temperature phases, the numerical data indicate a LRO for PD but a quasi-LRO in the KT-like region.

A similar phase diagram is obtained at $J=2$, as shown in Fig.~\ref{fig:ndiag}(b).
In this case also, the PD phase emerges in the intermediate-temperature region.
However, in contrast to the case with $J=1$ where PD is found widely above the FR state as well as PS, the PD phase dominantly appears above the PS region between the stripe and FR states.

\begin{figure}
   \includegraphics[width=0.8\linewidth]{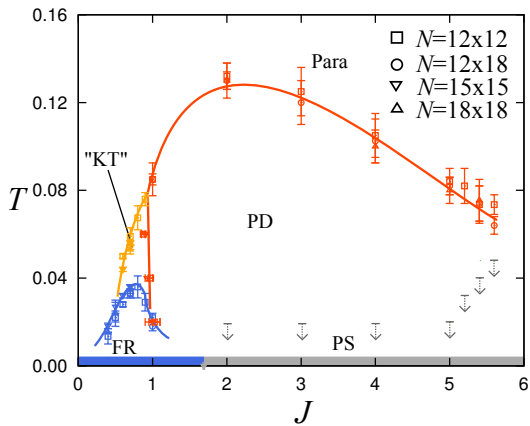}
   \caption{(Color online).
   Phase diagram of the model~(\ref{eq:H}) at $n=1/3$ while varying $J$.
   The notations are common to those in Fig.~\ref{fig:ndiag}.
   The boundary between PD and PS is difficult to determine by MC calculations, and supposed to be located at lower temperature than indicated by the gray arrows.
   }
   \label{fig:jdiag}
\end{figure}

We also investigated the phase diagram of the model in Eq.~(\ref{eq:H}) while varying $J$.
Figure~\ref{fig:jdiag} shows the numerically obtained phase diagram at $n=1/3$.
The result shows that the PD state is stable in a wide range of $0.8 \lesssim J \lesssim 5.6$.
The transition temperature first rapidly increases as increasing $J$, while it turns to a gradual decrease after showing a peak at $J\sim 2$.

An important observation in this constant-$n$ phase diagram is that the PD state does not survive down to $J \to 0$, and it is taken over by the KT-like and FR phases in the small $J$ region.
The absence of PD state in the $J\rightarrow 0$ limit implies that the RKKY interaction in the second-order perturbation theory is insufficient in stabilizing the PD state.
Moreover, the emergence of PD for $J > J_c \ne 0$ is consistently understood within the Slater mechanism discussed in Sec.~\ref{sec:mft}; the MC result of $J_c\sim 0.8$ is in good accordance with the mean-field argument of the critical value $\Delta_c = 2/3$.
The result clearly indicates that a non-perturbative effect of itinerant electrons plays a crucial role in stabilizing the PD state.

In the PD region in Fig.~\ref{fig:jdiag}, our MC data do not show clear sign of further transition while decreasing temperature before the MC calculations become unstable.
In the low temperature region, however, it becomes difficult to determine the chemical potential $\mu$ for $n=1/3$.
The lowest temperature of MC calculations are shown in the phase diagram by the gray downward arrows. 
On the other hand, the analysis of the ground state indicates that the ground state for $J \lesssim 1.68$ is the FR state, while the region for $J \gtrsim 1.68$ is PS between the stripe and FR states.
In addition, we observe the PS instability by carefully investigating the change of $n$ as a function of $\mu$ at $J=5.4$ (see also Appendix~\ref{sec:pseparation}). 
From these facts, we conclude that the PD for $J \gtrsim 1.68$ is taken over by PS between the stripe and FR states.
Since it is tedious to determine the PS boundary from $\mu$-$n$ plot for all the values of $J$, we merely plot the lowest temperature we reached in our constant-$n$ calculations as the upper limit of temperature for the PS instability.

\subsection{
Partial disorder
\label{sec:pd}
}

\begin{figure*}
   \includegraphics[width=\linewidth]{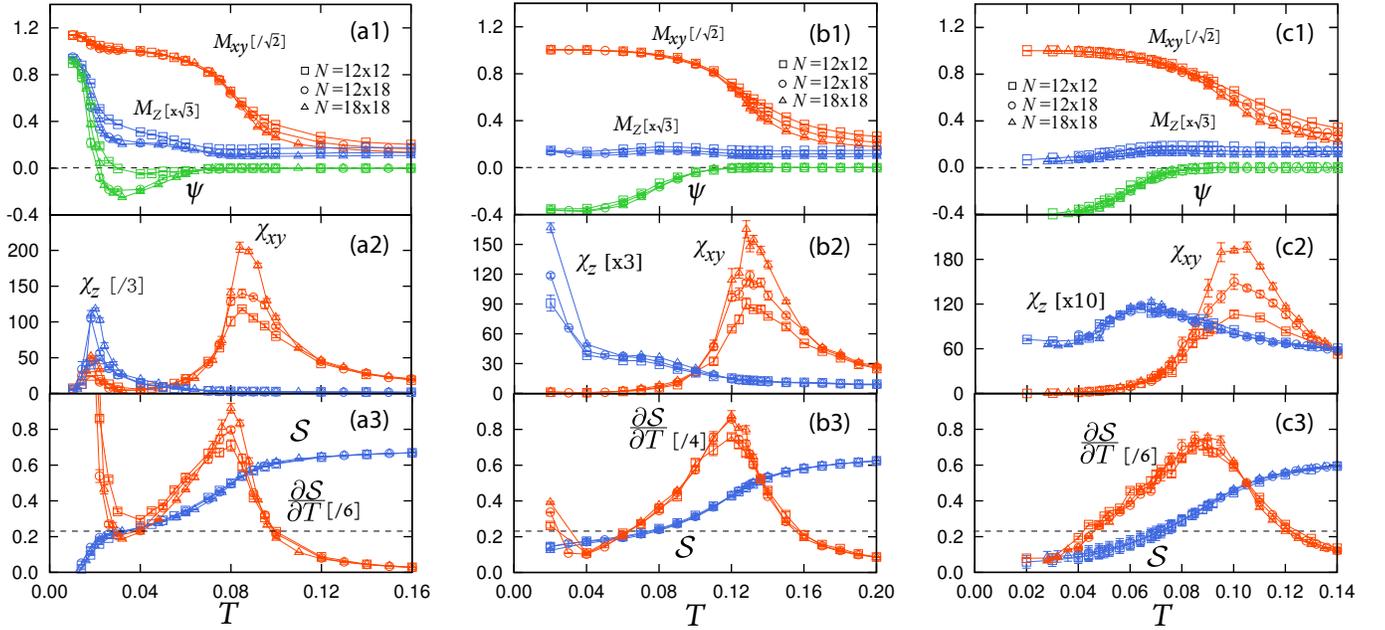}
   \caption{(Color online).
   MC results for (a1)-(c1) $M_{xy}$, $M_z$, and $\psi$, (a2)-(c2) $\chi_{xy}$ and $\chi_z$, and (a3)-(c3) $\cal{S}$ and its temperature derivative $\partial {\cal S}/\partial T$ at $n=1/3$;
   (a1)-(a3) $J=1$, (b1)-(b3) $J=2$, and (c1)-(c3) $J=4$.
   The calculations were done for the system sizes $N=12\times 12$, $12\times 18$, and $18\times18$.
   $\cal S$ is calculated from numerical integration of $\partial {\cal S}/\partial T$ by assuming ${\cal S}(T=1)=\log2$.
   }
   \label{fig:mcpd}
\end{figure*}

Here, we present the details of MC data for identifying the PD state.
Figure~\ref{fig:mcpd} shows $T$ dependences of MC results for different $J$ at $n=1/3$.
To fix $n$, we tuned $\mu$ for each temperature; the errors for $n$ at each temperature are controlled within 0.001.
Figure~\ref{fig:mcpd}(a1) is the result for the pseudomoments $M_{xy}$ and $M_z$ at $J=1$ [see the definitions in Eqs.~(\ref{eq:Mxy}) and (\ref{eq:Mz}), respectively].
$M_{xy}$ shows two anomalies while decreasing temperature at $T_c^{\rm (PD)} = 0.086(4)$ and $T_c^{\rm (FR)} = 0.019(2)$.
The critical temperatures are determined by the peaks of the susceptibilities, $\chi_{xy}$, and $\chi_z$, as mentioned below.
At $T_c^{\rm (PD)}$, $M_{xy}$ rapidly increases and approaches $\sqrt{2}$ at lower temperature. 
In addition, it shows a kink at $T_c^{\rm (FR)}$ and further increase to $8/3$ at lower temperature.
Meanwhile, $M_z$ shows no anomaly at $T_c^{\rm (PD)}$, while it shows a rapid increase to $1/\sqrt3$ at $T_c^{\rm (FR)}$.
Correspondingly, $\chi_{xy}$ and $\chi_z$ in Fig.~\ref{fig:mcpd}(a2) also show divergent peaks increasing with the system size;
peaks of $\chi_{xy}$ appear at both $T_c^{\rm (PD)}$ and $T_c^{\rm (FR)}$, while $\chi_z$ shows a peak only at $T_c^{\rm (FR)}$.
These results signal the presence of two successive phase transitions at $T_c^{\rm (PD)} = 0.086(4)$ and $T_c^{\rm (FR)} = 0.019(2)$.
The error bars are estimated by the range of temperature where the standard deviation of the MC data exceeds the difference of expectation value from the peak value.
The transition temperatures and error bars shown in Figs.~\ref{fig:ndiag} and \ref{fig:jdiag} are given by this criterion.
Meanwhile, most of the calculations in Fig.~\ref{fig:ndiag} were done by fixing $\mu$ instead of $n$.
Hence, we also give the error bars in terms of $n$, as $n$ changes with $T$ in a fixed $\mu$ calculation.

To determine the nature of low temperature phases at $n=1/3$, we also computed the azimuth parameter $\psi$ [Eq.~(\ref{eq:psi})] shown in Fig.~\ref{fig:mcpd}(a1).
While increasing the system sizes, $\psi$ apparently deviates from zero to a negative value below $T_c^{\rm (PD)}$, indicating that the intermediate phase for $T_c^{\rm (FR)} < T < T_c^{\rm (PD)}$ has a PD type order.
On the other hand, $\psi$ shows a sign change at $T_c^{\rm (FR)}$, and rapidly increases to $\psi=1$ at lower temperature.
This is a signature of the FR transition, which will be discussed in detail in Sec.~\ref{sec:stripe_ferri}.

The emergence of PD is also seen in the results for the spin entropy $\cal{S}$ and its temperature derivative [Eqs.~(\ref{eq:Sint}) and (\ref{eq:delSdelT}), respectively], as shown in Fig.~\ref{fig:mcpd}(a3).
In the intermediate-temperature region for $T_c^{\rm (FR)} < T < T_c^{\rm (PD)}$, $\cal{S}$ appears to approach $\frac13\log 2$ as decreasing temperature, which is the value expected for the ideal PD state where one out of three spins in the magnetic unit cell remains paramagnetic.
The remaining entropy is released rapidly at $T_c^{\rm (FR)}$ and ${\cal S} \to 0$ at lower temperature due to the ordering of paramagnetic spins in the FR state.

Similar phase transitions to the PD state are observed in the wide range of $J$, as shown in Figs.~\ref{fig:mcpd}(b) and \ref{fig:mcpd}(c) at $J=2$ and $J=4$, respectively.
In these results, however, we could not confirm the presence of another phase transition at a lower temperature in the range of temperature we calculated, in contrast to the FR transition found in the case of $J=1$.
As the PD state retains a finite $\cal{S}$, it is unlikely that this phase survives to $T\rightarrow0$.
Hence, it is presumably taken over by other ordered phases or PS at a lower temperature. 
As shown in Fig.~\ref{fig:jdiag}, the ground state is deduced to be PS for the values of $J$ in Figs.~\ref{fig:mcpd}(b)  and \ref{fig:mcpd}(c).
We, therefore, expect that the PD state is taken over by PS below $T=0.02$ for $J \gtrsim 2$. 
The situation is indicated by the gray arrows in the phase diagram in Fig.~\ref{fig:jdiag}, as discussed in Sec.~\ref{sec:pdiag}.

\begin{figure}
   \includegraphics[width=0.8\linewidth]{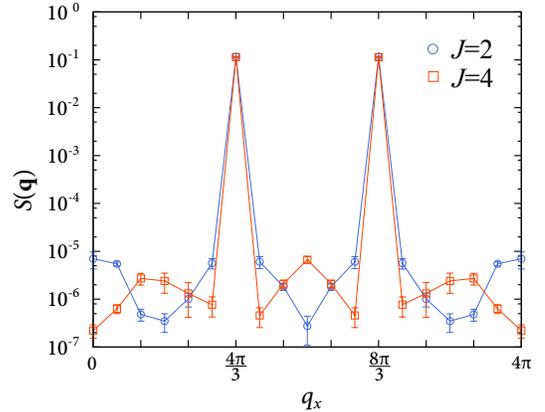}
   \caption{(Color online).
   MC results for $S({\bf q})$ along the ${\bf q}=(q_x,0)$ line at $T=0.02$.
   The calculations were done for the system size $N=18\times18$.
   }
   \label{fig:sq}
\end{figure}

Another point to be noted is the systematic change in $\cal{S}$ in the PD state by changing $J$.
While the result at $J=1$ appears to show plateau like behavior at ${\cal S}\sim \frac13 \log2$, the plateau value of ${\cal S}$ in the PD state decreases while increasing $J$, as shown in Figs.~\ref{fig:mcpd}(a3), \ref{fig:mcpd}(b3), and \ref{fig:mcpd}(c3).
The decrease in ${\cal S}$ is presumably attributed to the development of spatial correlations between paramagnetic sites in the PD state; 
the ideal value ${\cal S} = \frac13 \log2$ is for completely uncorrelated paramagnetic spins, and correlations between them reduces the entropy.
Such development of correlatins are observed in the spin structure factor $S({\bf q})$ defined in Eq.~(\ref{eq:Sq}).
Figure~\ref{fig:sq} shows a profile of $S({\bf q})$ calculated by MC simulation at $T=0.02$.
The peaks at ${\bf q}=(4\pi/3,0)$ and $(8\pi/3,0)$ indicates that the system is in a three-sublattice ordered phase, while the absence of a sharp peak at ${\bf q}=(0,0)$ indicates that there is no net magnetic moment; the result is consistent with PD order.
When comparing the results at $J=2$ and $J=4$, the peak corresponding to the three-sublattice order gets sharper for $J=4$, while the height of the peak of $S({\bf q})$ is almost the same. 
This indicates that the PD order at $J=2$ shows more spin fluctuations than that at $J=4$, consistent with the trend of the plateau value of ${\cal S}$.

\begin{figure}
   \includegraphics[width=0.8\linewidth]{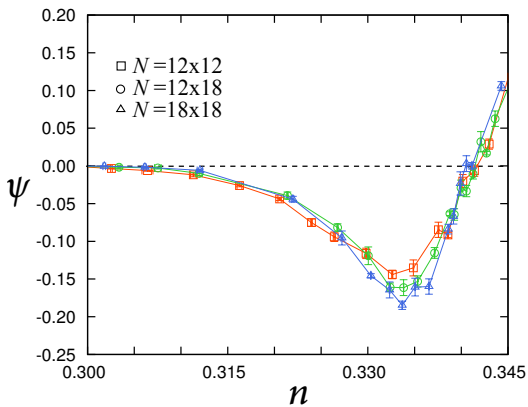}
   \caption{(Color online).
   MC results for $\psi$ while varying $n$ at $T=0.08$ and $J=2$.
   The calculations were done for the system sizes $N=12\times 12$, $12\times 18$, and $18\times18$.
   }
   \label{fig:conT}
\end{figure}

Thus far, we showed the results at $n=1/3$.
Next, we show how the PD evolves while changing $n$.
Figure~\ref{fig:conT} shows the MC result of $\psi$ as a function of $n$ at $T=0.08$ and $J=2$.
$\psi$ becomes negative around $n=1/3$ and takes the lowest value at $n\simeq1/3$. 
The data indicate that $\psi$ is almost system size independent or rather slightly decreases as the system size increases in the finite range of $n$ around $n=1/3$. 
Hence, the PD state is stabilized not only at $n=1/3$ but for a finite range of $0.31 \lesssim n \lesssim 0.34$ in the thermodynamic limit. 
The range well agrees with that for the PD phase estimated from the peak of susceptibilities shown in Fig.~\ref{fig:ndiag}(b).

With regard to the order of the PD transition, the PD transition in our MC results appears to be continuous, as shown in Fig.~\ref{fig:mcpd}. 
However, it needs careful consideration, as we will discuss here.
It is known that the Ising model on a triangular lattice with AF NN interactions is effectively described by a six-state model, in which the low-energy states with three up-up-down and three up-down-down configurations in the three-site unit cell are described by six-state variables. 
The PD state in our model also retains six low-energy states with different up-down-paramagnetic configurations, and hence, the transition to PD is expected to be classified in the framework of six-state models.
However, from the argument of duality properties, it is prohibited that the six-state models exhibit a single second-order transition for changing temperature~\cite{Cardy1980}.
For instance, a two-dimensional six-state clock model shows two KT transitions at finite temperature, without exhibiting true LRO for $T\ne0$~\cite{Jose1977,Challa1986}.
On the other hand, a six-state Potts model shows a weak first order transition to LRO, in which the correlation length reaches the order of 1000 sites at the critical point~\cite{Buffenoir1993}.
In our PD case, the apparently second-order transition at $T_c^{\rm (PD)}$ is not expected to be a single one, but is always followed by another transition to FR or PS at a lower temperature. 
This appears not to violate the general argument for the six-state models, although it is not clear to what extent the argument applies, as the electronic PS never takes place in the localized spin models.
Hence, the PD transition can be of second order, as indicated in our numerical results. 
Of course, we cannot exclude the possibility of a weak first order transition, similar to that of the Potts model.
In this case, due to a long correlation length at the critical temperature, the system sizes used in our calculations are likely to be insufficient to distinguish the first order transition from second order one.

\subsection{
Other magnetic orders
\label{sec:stripe_ferri}
}

\begin{figure}
   \includegraphics[width=0.9\linewidth]{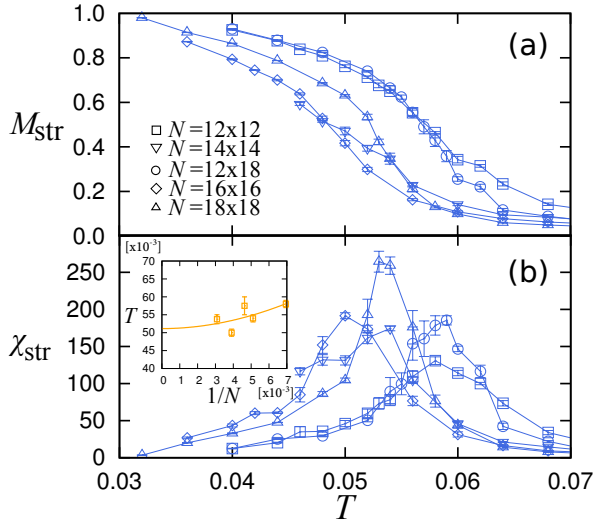}
   \caption{(Color online).
   MC results for (a) $M_{\rm str}$ and (b) its susceptibility $\chi_{\rm str}$ at $J=2$ and $n=0.27$.
   The inset in (b) shows $T_c^{\rm (str)}$ for different sizes and the solid line is the extrapolation which gives $T_c^{\rm (str)}=0.051(13)$.
   The calculations were done for the system sizes $N=12\times 12$, $14\times 14$, $12\times 18$, $16\times 16$, and $18\times18$.
   }
   \label{fig:mcstripe}
\end{figure}

Figure~\ref{fig:mcstripe} presents the results for the relatively low filling where the stripe order is stabilized at low temperature.
Figure~\ref{fig:mcstripe}(a) shows the order parameter for the stripe order, $M_{\rm str}$ [Eq.~(\ref{eq:Mstr})], and Fig.~\ref{fig:mcstripe}(b) shows the corresponding susceptibility $\chi_{\rm str}$ at $J=2$ and $n=0.27$.
A phase transition to the stripe phase is signaled by a rapid increase of $M_{{\rm str}}$ and corresponding peak of $\chi_{\rm str}$;
we determine the transition temperature $T_c^{\rm (str)}$ by the peak temperature of $\chi_{\rm str}$ for each system size, and plot them in the phase diagram in Fig.~\ref{fig:ndiag}(a). 
The error bars are estimated in a similar manner to the case of $T^{\rm (PD)}_c$ and $T^{\rm (FR)}_c$.
We also show the system-size extrapolation of $T_c^{\rm (str)}$ in the inset of Fig.~\ref{fig:mcstripe}(b).
Although the data are rather scattered, we fit them by $f(N) = a + b/N^c$ with fitting parameters $a$, $b$, and $c$. 
The extrapolation clearly shows that the phase transition takes place at a finite temperature, as expected for the two-dimensional Ising order.

The stripe ordered phase is a peculiar magnetic state, in which the sixfold rotational symmetry of the lattice is spontaneously broken and reduced to twofold.
Due to the symmetry breaking, the transport property is expected to show strong spatial anisotropy; e.g., the longitudinal conductivity will be large in the direction along the stripes, while suppressed in the perpendicular direction.
This is an interesting topic on the control of transport by magnetism and vice versa.

\begin{figure}
   \includegraphics[width=0.76\linewidth]{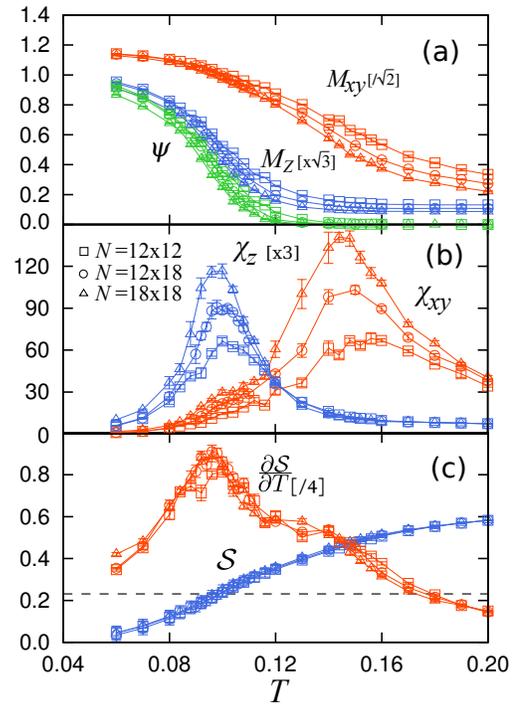}
   \caption{(Color online).
   MC results for (a) $M_{xy}$, $M_z$, and $\psi$, (b) $\chi_{xy}$ and $\chi_z$, and (c) $\cal{S}$ and its temperature derivative $\partial {\cal S}/\partial T$ at $n=0.38$ and $J=2$.
   The calculations were done for the system sizes $N=12\times 12$, $12\times 18$, and $18\times18$.
   }
   \label{fig:mcferri}
\end{figure}

Figure~\ref{fig:mcferri} shows the results for the relatively high filling where the low temperature phase is FR, at $n=0.38$ and $J=2$.
The data indicate two successive transitions signaled by the peaks in $\chi_{xy}$ and $\chi_{z}$ at different temperature.
The peak of $\chi_z$ corresponding to the increase of $M_z$ signals the phase transition to the FR phase at $T_c^{\rm (FR)} = 0.098(4)$.
At the same time, $\psi$ becomes finite below $T_c^{\rm (FR)}$, and approaches 1, as expected for the FR ordering.
Similar behavior was observed at $T_c^{\rm (FR)} = 0.019(2)$ in Figs.~\ref{fig:mcpd}(a1) and \ref{fig:mcpd}(a2).
On the other hand, at a higher $T_{\rm KT} = 0.146(4)$, only $M_{xy}$ changes rapidly, and correspondingly, $\chi_{xy}$ shows a peak.
$M_{xy}$, however, shows a noticeable system-size dependence even below $T_{\rm KT}$, in contrast with the results below $T_c^{\rm (PD)}$. 
Similar behavior was observed in the KT transition in Ising spin systems~\cite{Takayama1983,Fujiki1984}.

\begin{figure}
   \includegraphics[width=0.8\linewidth]{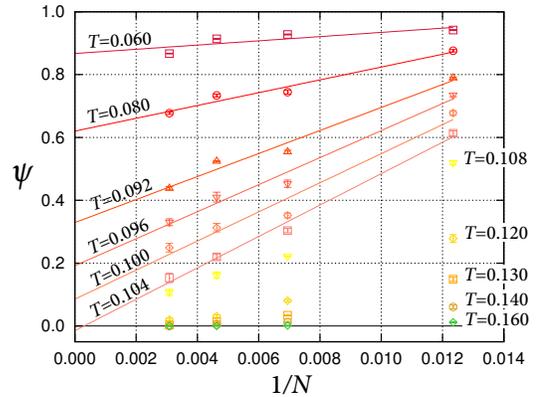}
   \caption{(Color online).
   Extrapolation of $\psi$ to $N\to\infty$ at different temperatures.
   The solid lines for $T\le 0.104$ is the linear fitting of data.
   }
   \label{fig:psifit}
\end{figure}

On the other hand, $\psi$ does not show an anomaly at $T_{\rm KT}$, while it shows a sharp rise around $T_{c}^{{\rm (FR)}}$, as shown in Fig.~\ref{fig:mcferri}(a).
The value of $\psi$ extrapolated to large $N$ converges to zero in the intermediate-temperature range.
Figure~\ref{fig:psifit} shows the extrapolation of $\psi$ for $N\to \infty$.
The results indicate that, $\psi$ remains to be zero at $N\to\infty$ for $T\gtrsim0.104$, which is far below $T_{\rm KT}=0.146(4)$.
On the other hand, the extrapolated value becomes finite for $T\lesssim 0.104$, reflecting the FR order; the transition temperature is estimated as $\tilde{T}_c^{\rm (FR)} = 0.102(2)$, which is in accordance with $T_c^{\rm (FR)} = 0.098(4)$.

The results above indicate that there is no sixfold symmetry breaking in $M_{xy}$ at $T_{\rm KT}$, as seen in the KT phase in the Ising spin models~\cite{Takayama1983}.
Hence, we consider that the higher-temperature transition at $T_{\rm KT}$ is of KT type.
Namely, the system exhibits two successive transitions from the paramagnetic phase to the KT-like phase at $T_{\rm KT}$, and the KT-like phase to the low-temperature FR phase at $T_{c}^{{\rm (FR)}}$. 
Here, we call the intermediate-temperature phase the KT-like phase, as it is difficult to confirm either the KT universality class by critical behavior or the quasi-LRO behavior within the system sizes we reached, as seen below.

\begin{figure}
   \includegraphics[width=0.8\linewidth]{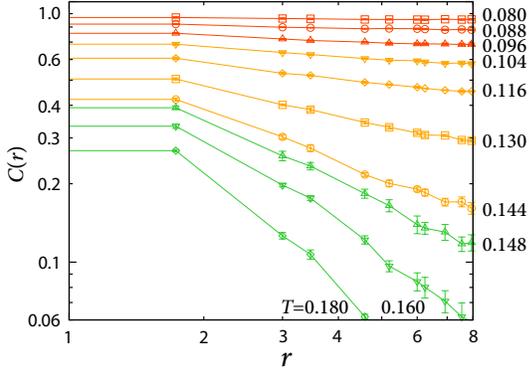}
   \caption{(Color online).
   MC results for the real-space spin correlation function $C(r)$ at $J=2$ and $n=0.38$.
   The results are shown only for the sites with $C(r) > 0$.
   The calculations were done for the system size $N=18\times18$.
   }
   \label{fig:mcsk}
\end{figure}

The signature of two successive transitions is also observed in the real-space spin correlation function $C(r)$.
Here $C(r)$ is the averaged correlations between the Ising spins in distance $r$, defined by 
\begin{eqnarray}
C(r) = \sum_{i,j} \frac{1}{N_{\rm p}(r)} \langle S_i S_j \rangle  \delta(|{\bf r}_{ij}| - r),
\end{eqnarray}
where $N_{\rm p}(r) =\sum_{i,j} \delta(|{\bf r}_{ij}| - r)$ is the number of spin pairs with distance $r$, and $\delta(x)$ is the delta function.
The MC data while varying temperature are shown in Fig.~\ref{fig:mcsk}. 
Although the results are not conclusive due to the limitation on accessible system sizes, they appear to be consistent with the two transitions discussed above.
For $T \lesssim T_{c}^{{\rm (FR)}} = 0.098(4)$, the spin correlation appears to approach constant for large distance, well corresponding to the FR LRO developed in this low temperature region.
On the other hand, for $T \gtrsim T_{\rm KT} = 0.146(4)$, it becomes concave downward with a steep decrease with respect to the distance, which reflects an exponential decay in the high temperature paramagnetic state.
In the intermediate region for $T_c^{\rm (FR)} \lesssim T \lesssim T_{\rm KT}$, the spin correlation also decays with increasing distance.
The decay, however, is much slower and appears to obey an asymptotic power law, which is characteristic to the quasi-LRO in the KT state.
In principle, the critical exponents can be estimated from the asymptotic power-law behavior, but it is difficult to be conclusive in the current system sizes.

\section{
Electronic structure of partially disordered state
\label{sec:estruct}
}

In the previous section, we discussed the thermodynamic behavior of the localized spin degree of freedom, with emphasis on the emergence of peculiar PD state.
In this section, we focus on the behavior in the charge degree of freedom of itinerant electrons in the PD phase. 

\begin{figure}
   \includegraphics[width=0.80\linewidth]{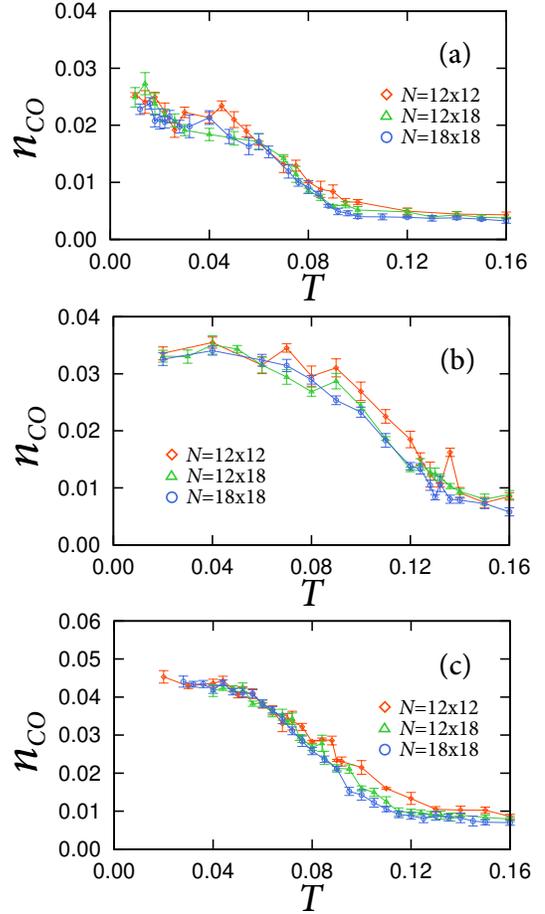}
   \caption{(Color online).
   MC results for $n_{\rm CO}$ at ${\bf q}=(2\pi/3,-2\pi/3)$ at $n=1/3$ and (a) $J=1$, (b) $J=2$, and (c) $J=4$.
   The calculations were done for the system sizes $N=12\times 12$, $12\times 18$, and $18\times18$.
   }
   \label{fig:mcnq}
\end{figure}

Figure~\ref{fig:mcnq} shows temperature dependence of the charge modulation $n_{\rm CO}$ [Eq.~(\ref{eq:n_CO})] at $n=1/3$ for different $J$. 
Figure~\ref{fig:mcnq}(a) is the result at $J=1$ for different system sizes.
The result shows an increase of $n_{\rm CO}$ below $T \simeq T_c^{\rm (PD)} =0.086(4)$, indicating that the PD state is accompanied by charge modulation with period three.
Similar onsets of charge modulation at $T_c^{\rm (PD)}$ are observed for larger $J$, as shown in Figs.~\ref{fig:mcnq}(b) and \ref{fig:mcnq}(c);
the amplitude of the modulation in the PD phase increases monotonically as $J$ increases.
The magnitude of the charge modulation is in the same order compared to the mean-field result in Fig.~\ref{fig:gap}), while the growth is considerably suppressed by a factor of two to four.

\begin{figure}
   \includegraphics[width=0.8\linewidth]{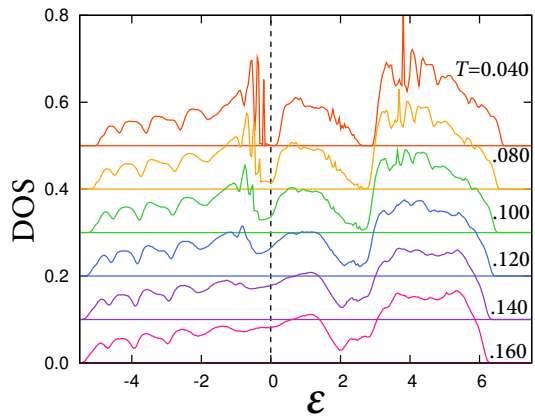}
   \caption{(Color online).
   MC results for DOS of itinerant electrons at $n=1/3$ and $J=2$ for $N=18\times 18$.
   The Fermi level is set at $\varepsilon = 0$.
   The statistical errors are comparable to the width of the lines.
   }
   \label{fig:mcdos}
\end{figure}

We next look into the electronic density of states (DOS) at different temperature. 
Figure~\ref{fig:mcdos} shows the results for DOS while varying temperature at $J=2$ and $n=1/3$.
The Fermi level is set at $\varepsilon=0$.
Here, DOS was calculated by counting the number of energy eigenvalues as the histogram with the energy interval of 0.0375.
In the paramagnetic region for $T \gtrsim T_c^{\rm (PD)} = 0.130(4)$, DOS is featureless near the Fermi level.
On the other hand, below $T_c^{\rm (PD)}$, an energy gap develops at the Fermi level for $n=1/3$.
The result shows that the PD state is an insulator, which supports the scenario that PD is stabilized by the Slater mechanism described in Sec.~\ref{sec:mft}.
Similarly to the charge modulation, the energy gap in the MC results is largely suppressed compared to that obtained by the mean-field analysis in Fig.~\ref{fig:gap}.
This appears to show the importance of appropriately taking into account of thermal fluctuations.

\section{\label{sec:summary}
Summary
}

To summarize, by a combined analysis of the mean-field type calculation and Monte Carlo simulation, we have investigated the origin of the partial disorder in the Ising-spin Kondo lattice model in a two-dimensional triangular lattice. 
In the mean-field type calculation, we have clarified that a local magnetic field of the partial disorder type induces a metal-insulator transition at 1/3 filling at a critical value of the field. 
The result suggests that the three-sublattice partial disorder can give rise to an energy gap, and therefore, it has a chance to be stabilized through the Slater mechanism. 
On the other hand, in the Monte Carlo simulation, we have provided convincing numerical results on the emergence of partial disorder at finite temperatures where the stripe phase and the ferrimagnetic order compete with each other.
The Monte Carlo result shows that the partially disordered state appears above a nonzero value of the spin-charge coupling, and that it is insulating and accompanied by charge disproportionation.
The nonzero critical value of the spin-charge coupling and the opening of the charge gap are both qualitatively consistent with the mean-field analysis.
The results indicate that the partial disorder is stabilized by the Slater mechanism which is characteristic to itinerant magnets.
Our results not only clarify the new mechanism of partial disorder in two dimensions but also pave the way for understanding of the interesting physics related to the peculiar coexistence of magnetic order and paramagnetic moments in itinerant electron systems.

\begin{figure}
   \includegraphics[width=0.8\linewidth]{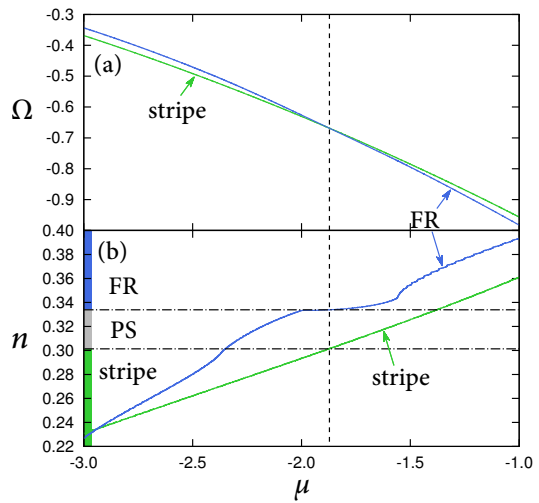}
   \caption{
   (Color online).
   (a) The grand potetial $\Omega$ and (b) electron filling $n$ with respect to the chemical potential $\mu$, numerically calculated by exactly diagonalizing the one-body Hamiltonian for itinerant electrons.
   The results are obtained at $J=2$ with $N_s=24\times24$ site superlattice of $N=12\times 12$ site unit cells.
   The strip at the left side of (b) shows the ground state at the corresponding filling.
   }
   \label{fig:nvc}
\end{figure}

An interesting extension of the current work would be to consider the effect of quantum fluctuation of localized spins.
In our result, the partial disorder remains stable down to very low temperature, implying that the paramagnetic spins are largely fluctuating and sensitive to perturbations at low temperatures.
Hence, an interesting possibility is that, by including quantum fluctuations, the partial disorder is further stabilized and remains stable even in the ground state.
Indeed, a similar partial disorder was found in the ground state of the Kondo lattice model with quantum spins at half filling~\cite{Motome2010}. 
Therefore, it is intriguing to examine the effect of quantum fluctuations on the present model with Ising spins. 
However, it is not straightforwardly calculated by the present Monte Carlo method.
The interesting problem is left for future study.

\begin{acknowledgements}
The authors are grateful to G.-W. Chern, H. Kawamura, M. Matsuda, S. Miyashita, and H. Yoshida for fruitful discussions.
The authors also thank S. Hayami and T. Misawa for helpful comments.
Part of the calculations were performed on the Supercomputer Center, Insitute for Solid State Physics,
University of Tokyo. H.I. is supported by Grant-in-Aid for JSPS Fellows.
This research was supported by KAKENHI (No.19052008, 21340090, 22540372, and 24340076), Global COE Program ``the Physical Sciences Frontier", the Strategic Programs for Innovative Research (SPIRE), MEXT, and the Computational Materials Science Initiative (CMSI), Japan.
\end{acknowledgements}

\appendix
\section{\label{sec:pseparation}
Phase separation
}

In this appendix, we present how to identify the PS region.
First, we show the method we used to determine the ground state phase diagram shown in Figs.~\ref{fig:ndiag} and \ref{fig:jdiag}.
The ground state is obtained by variational calculations, i.e., by comparing the grand potential per site, $\Omega = \langle H \rangle/N - \mu n$, where $\mu$ is the chemical potential and $n$ is the electron filling. 
Here, we compare $\Omega$ calculated for the magnetically ordered states, stripe and FR, which appear in the MC simulation at low temperature in the present parameter regions.
The procedure is shown in Fig.~\ref{fig:nvc} at $J=2$.
Figures~\ref{fig:nvc}(a) and \ref{fig:nvc}(b) show the results of $\Omega$ and $n$, respectively, calculated for stripe and FR orders.  
For $\mu \lesssim -1.87$ ($\mu \gtrsim -1.87$), $\Omega$ for the stripe order is lower (higher) than that for the FR order, indicating that the stripe (FR) state is the ground state in this region. 
At the critical value of $\mu \simeq -1.87$, the electron filling for the two states take different values, $n \simeq 0.301$ in the stripe state and $n \simeq 0.334$ in the FR state, as shown in Fig.~\ref{fig:nvc}(b). 
This indicates that $n$ changes discontinuously from $n \simeq 0.301$ to $n \simeq 0.334$ at the transition between the stripe and FR states. 
In other words, the system is unstable in the region of $0.301 \lesssim n \lesssim 0.334$ against PS between the two states; the range of $n$ is identified as the electronic PS. 
The PS regions in Fig.~\ref{fig:ndiag} are determined in this manner. 
Meanwhile, the PS region at $n=1/3$ in Fig.~\ref{fig:jdiag} is identified by the similar calculations by changing $J$.

\begin{figure}
   \includegraphics[width=0.8\linewidth]{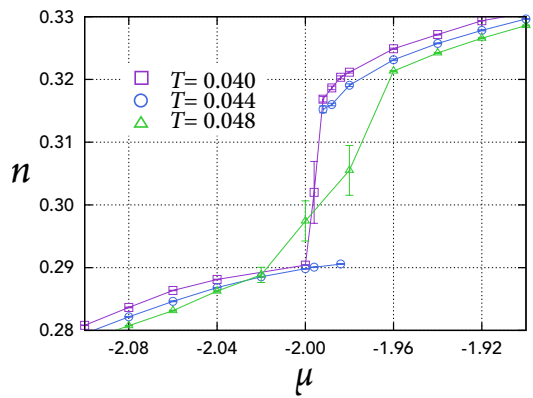}
   \caption{(Color online).
   MC results for $n$ as a function of $\mu$ at different temperature.
   The results are for $J=2$ and $N=12\times 12$.
   }
   \label{fig:muvsn}
\end{figure}

Next, we describe how the PS region is determined at finite temperature in the MC calculation.
In the MC simulation using the grand canonical ensemble, PS is characterized by a sudden jump of $n$ while sweeping $\mu$.
Figure~\ref{fig:muvsn} shows a typical MC result for $n$ as a function of $\mu$.
The result at $T=0.048$ shows a smooth change of $n$ in the entire region of $\mu$ in the figure.
On the other hand, the results at $T=0.040$ and $0.044$ show a sudden change from $n\sim0.290$ to $0.315$ at $\mu\sim -1.996$.
We roughly estimate the PS region by the values of $n$ at the both ends of the jump.
The results are plotted in the phase diagrams in Fig.~\ref{fig:ndiag}. 
The range of PS slightly depends on the system size, and hence, we plot the threshold values of $n$ for each system size in the phase diagram.

\end{document}